\documentstyle[12pt,psfig]{article}
\newcommand{\beq}{\begin{equation}}
\newcommand{\eeq}{\end{equation}}
\newcommand{\beqa}{\begin{eqnarray}}
\newcommand{\eeqa}{\end{eqnarray}}
\newcommand{\ba}{\begin{array}}
\newcommand{\ea}{\end{array}}

\begin{document}

\begin{center}
{\Large \bf Bose-Einstein Condensation \\
in Exotic Trapping Potentials} 
\vskip 0.5cm 
{Luca Salasnich}
\vskip 0.5cm 
{Istituto Nazionale per la Fisica della Materia, Unit\`a di Milano,\\
Dipartimento di Fisica, Universit\`a di Milano, \\
Via Celoria 16, 20133 Milano, Italy} 
\end{center}

\vskip 0.5cm 

\begin{center}
{\bf Abstract}
\end{center} 
We discuss thermal and dynamical properties 
of Bose condensates confined by an external potential. 
First we analyze the Bose-Einstein transition temperature 
for an ideal Bose gas in a generic power-law potential and 
d-dimensional space. Then we investigate the effect of the 
shape of the trapping potential on the 
properties of a weakly-interacting 
Bose condensate. We show that using exotic trapping potentials 
the condensate can exhibit interesting coherent quantum phenomena, 
like superfluidity and tunneling. 
In particular, we consider toroidal and double-well 
potentials. The theoretical results are compared with 
recent experiments. 

\vskip 0.5cm 

\section{Introduction}

In 1995 the Bose-Einstein condensation (BEC), 
i.e. the macroscopic occupation 
of the lowest single-particle state of a system of bosons, 
has been experimentally achieved with clouds of confined alkali-metal 
atoms at ultra-low temperature (about $100$ nK). 
In that year, three different groups 
(at JILA with $^{87}$Rb atoms [1], 
at MIT with $^{23}$Na atoms [2], and at Rice University 
with $^{7}$Li atoms [3]) have obtained 
the BEC by using the same technique:  
a laser cooling and confinement in a magnetic trap 
and an evaporative cooling. 
In 1997 the Nobel Prize in Physics has been given for 
the development of methods to cool and 
trap atoms with laser light [4]. 
Nowadays more than twenty experimental groups have achieved BEC by using 
different geometries of the confining trap and atomic species. 
\par 
The BEC phase transition has been predicted for a homogeneous 
ideal gas of Bosons by Einstein [5] on the basis of 
a paper of Bose [6]. The idea of Bose condensate 
has been used by London to describe the superfluid 
behavior of $^{4}$He [7]. $^{4}$He is a strongly-interacting 
system and the condensed fraction does not excedes $10$\%. 
Instead, in the dilute alkali-metal atoms of recent experiments,  
the condensed fraction can be more than $90$\%. 
Thus, dilute vapors are ideal systems to 
investigate thermal and dynamical properties 
of the Bose condensate and the role of the trapping potential. 
\par 
In this review paper we discuss some recent theoretical 
results we have obtained for Bose gases confined 
in potentials with exotic shapes, like power-law, 
toroidal and double-well potentials. 
Exotic potentials can be used in future experiments 
to find signatures of new coherent matter-wave phenomena. 

\section{Confined Bose gas and BEC temperature} 

In this section we investigate a confined ideal quantum gas of 
Bosons. By using the semiclassical 
approximation, we derive analytical formulas for the BEC transition 
temperature. 
\par 
In the grand canonical ensemble of equilibrium 
statistical mechanics [8], 
the average number $N_{\alpha}$ of Bosons in the single-particle 
state $|\alpha\rangle$ with energy $\epsilon_{\alpha}$ is given by 
\beq 
N_{\alpha} = {1\over e^{\beta(\epsilon_{\alpha}-\mu)} - 1}  \; , 
\eeq 
where $\mu$ is the chemical potential and $\beta=1/(kT)$ 
with $k$ the Boltzmann constant and $T$ the absolute temperature. 
In general, given the single-particle function $N_{\alpha}$, 
the average total number $N$ of particles of the system reads 
\beq 
N= \sum_{\alpha} N_{\alpha}  \; . 
\eeq 
This condition fixes the chemical potential $\mu$. 
Thus, $\mu$ is a function of $\beta$ and $N$. 
$\mu$ cannot be higher than the lowest single-particle energy level 
$\epsilon_0$, i.e., it must be $\mu < \epsilon_0$. 
When $\mu$ approaches $\epsilon_0$ the function $N_0$ becomes 
very large and consequently $N_0$ becomes of the same order 
of $N$. The physical meaning is that the lowest 
single-particle state becomes macroscopically occupied and 
one has the so-called Bose-Einstein condensation (BEC). 
It is clear that if $\mu = \epsilon$ then $N_0$ 
diverges. It is a standard procedure to calculate the condensed 
fraction $N_0/N$ and also the BEC transition temperature $T_B$ 
by studying the non divergent quantity $N-N_0$ at $\mu =\epsilon_0$ 
as a function of the temperature. This procedure is particularly 
effective by using the semiclassical approximation. 
\par 
In the semiclassical approximation, the 
system is described by a continuum 
of states and, instead of $\epsilon_{\alpha}$, 
one uses the classical single-particle phase-space energy 
\beq
\epsilon({\bf r},{\bf p})=
{{\bf p}^2\over 2m} + U({\bf r}) \; , 
\eeq 
where ${\bf p}^2/(2m)$ is the kinetic energy 
and $U({\bf r})$ is the confining external potential. 
In this way one obtains 
the semiclassical single-particle phase-space distribution 
of Bosons 
\beq 
n({\bf r},{\bf p}) = 
{1\over e^{\beta(\epsilon({\bf r},{\bf p})-\mu)} - 1}  \; . 
\eeq 
Note that the accuracy of the semiclassical approximation is expected 
to be good if the number of particles is large 
and the energy level spacing is smaller then $kT$ [9]. 
\par
For the sake of generality, let us consider a d-dimensional space. 
Then the quantum elementary volume of the single-particle 
2d-dimensional phase-space is 
given by $(2\pi\hbar)^d$, where $\hbar$ is the Planck constant [9]. 
It follows that the average number $N$ of Bosons 
in the d-dimensional space can be written as 
\beq
N= \int {d^d{\bf r} \; d^d{\bf p} \over (2\pi \hbar )^d}
\; n({\bf r},{\bf p}) =
\int d^d{\bf r} \; n({\bf r}) \; ,
\eeq
where 
\beq
n({\bf r})=\int {d^d{\bf p} \over (2\pi \hbar )^d} 
n({\bf r},{\bf p}) =
{1\over \lambda^d} g_{d\over 2}
\left(e^{\beta(\mu -U({\bf r}))}\right) \; ,
\eeq 
is the spatial distribution, 
$\lambda = (2\pi \hbar^2\beta /m)^{1/2}$ is the thermal length, 
$$ 
g_{n}(z)= {1\over \Gamma(n)}\int_0^{\infty} dy 
{z e^{-y} y^{n-1} \over 1 - z e^{-y} } \; ,
$$
is the Bose function, and $\Gamma(n)$ is the factorial function. 
\par 
It is important to observe that, while Eq. (4) is divergent 
at $\mu = \epsilon({\bf r},{\bf p})$, Eq. (6) is not divergent 
at $\mu = U({\bf r})$ because $g_{d/2}(1)=\zeta(d/2)$, 
where $\zeta(x)$ is the Riemann zeta-function. 
It means that the integration over momenta 
removes the divergent contribtion coming from the 
ground-state. It follows that the Eq. (5) 
describes only the total number of non-condensed Bosons. 
Note that this number can also be written as 
\beq 
N=\int_0^{\infty} 
d\epsilon \; \rho(\epsilon ) \; 
{1\over e^{\beta(\epsilon - \mu)} - 1} \; , 
\eeq 
where $\rho (\epsilon )$ is the density of states. 
It can be obtained from the semiclassical formula 
\beq
\rho(\epsilon ) = \int {d^d{\bf r} \; d^d{\bf p}\over (2\pi\hbar)^d} 
\delta (\epsilon - \epsilon({\bf p},{\bf r})) = 
\left({m\over 2\pi \hbar^2}\right)^{d\over 2} 
{1\over \Gamma({d\over 2})} 
\int d^d{\bf r} \; 
\left( \epsilon - U({\bf r}) \right)^{(d-2)\over 2} \; . 
\eeq 
where $\delta(x)$ is the Dirac delta function and 
$\Gamma(n)$ is the factorial function. 
This result is the generalization 
of the formula for ideal homogeneous 
Bose gases in a box of volume $V$ [8]. 
It shows that, in the semiclassical limit, 
the non-homogeneous formula is obtained 
with the substitution $\mu \to \mu - U({\bf r})$, 
also called local density approximation. 

\subsection{BEC temperature for power-law potentials} 

In many experiments with alkali-metal atoms, the 
external trap can be accurately described by a harmonic potential. 
More generally, one can consider power-law potentials, given by  
\beq 
U({\bf r})= A \; r^n = {1\over 2} \left(
{\hbar \omega_0\over r_0^n}\right) \; r^n \; , 
\eeq
where $n$ is the power-law exponent and $A$ is the trap constant. 
Clearly, with $n=2$ one gets the harmonic potential. 
Here we have introduced an energy parameter $\hbar \omega_0$  
and a length parameter $r_0$, which can be chosen 
as $r_0=\sqrt{\hbar/(m\omega_0)}$. 
The power-law potential is interesting for studying 
the effects of adiabatic changes in the trap. 
The density of states of a quantum 
gas in the power-law potential 
can be calculated from Eq. (8) and reads 
\beq 
\rho(\epsilon ) = {2^{d\over n} \Gamma({d\over n}+1) 
\over 2^{d\over 2}\Gamma({d\over 2}+1) 
\Gamma({d\over 2} +{d\over n})} 
\left(\hbar\omega_0\right)^{2n\over d(n+2)} 
\epsilon^{{d(n+2)\over 2n}-1}  \; . 
\eeq 
\par 
As previously stated, Eq. (5) and (7) give the total 
number of Bosons only above critical temperature $T_B$. 
Below $T_B$, Eq. (5) and (7) describe 
the non-condensed number of particles $N-N_0$, where 
$N_0$ is the number of condensed particles. 
At the Bose transition temperature $T_B$, the chemical 
potential is $\mu=0$ and in this way one obtains 
the critical temperature. In the case of the power-law potential 
the critical temperature is 
\beq 
k T_B = c(d,n) \; \hbar \omega_0 \; N^{2n\over d(n+2)} \; , 
\eeq
where $c(d,n)$ is a numerical coefficient given by 
\beq
c(d,n)= 
\left[ {2^{d\over 2}\Gamma({d\over 2}+1) \over 2^{d\over n}
\Gamma({d\over n}+1) \zeta({d\over n} + {d\over 2}) } 
\right]^{2 n\over d(n+2)} 
\eeq
Below the critical temperature $T_B$, one has $N_0\neq 0$ 
and from Eq. (5) and (7) one gets the $T$ dependence 
of the condensed fraction 
$$
{N_0\over N} =1-\left({T\over T_B}\right)^{d(n+2)\over 2n} \; , 
$$ 
where $N$ is the number of Bosons in the gas. 
\par 
It is important to observe that from the previous formulas 
one easily derives the thermodynamic properties of quantum gases 
in harmonic traps and in a rigid box. In fact, by setting $n =2$ one gets 
the formulas for the Bose and Fermi gases in a harmonic trap 
(in the case of a anisotropic harmonic potential, $\omega_0$ 
is the geometric average of the frequencies of the trap). 
The results for a rigid box are instead obtained by letting 
${d\over n}\to 0$, where the density of particles 
per unit length is given by $N/\Omega_d$ and 
$\Omega_d= d\pi^{d\over 2}/\Gamma({d\over 2}+1)$ 
is the volume of the d-dimensional unit sphere. 
\par 
Finally, one notes that in the formula of the 
BEC transition temperature $T_B$ 
it appears the function $\zeta({d\over 2}+{d\over n})$. 
Because $\zeta(x)< \infty$ for $x>1$ but 
$\zeta(1)=\infty$, one easily deduces 
the following theorem [9]. 

\vskip 0.5cm 

{\bf Theorem}. Let us consider an ideal Bose gas 
in a power-law isotropic potential $U({\bf r})=A\; r^n$ 
with $r=|{\bf r}|=(\sum_{i=1}^d x_i^2)^{1/2}$. 
BEC is possible if and only if the following condition 
is satisfied  
$$
{d(n+2)\over 2n} > 1\; ,
$$
where $d$ is the space dimension and $n$ is the exponent of the 
confining power-law potential.  

\vskip 0.5cm 

This is a remarkable inequality. For example, 
for $d=2$ one finds the familiar result that 
there is no BEC in a homogeneous gas 
(${d\over n} \to 0$) but BEC is possible in a harmonic trap 
($n =2$). Moreover, one obtains that 
for $d=1$ BEC is possible with $1 < n < 2$. 

\section{Bose condensate in external potential} 

In this section we study in detail the static and dynamical 
properties of a trapped dilute Bose condensate at zero temperature. 
Note that, if the system of Bosons is dilute, at zero temperature the 
non-condensed fraction can be neglected [10]. 
\par
Let us consider a $N$-body quantum system with 
Hamiltonian ${\hat H}$. The exact time-dependent Schr\"odinger equation 
can be obtained by imposing the quantum least action 
principle to the action 
\beq 
S= \int dt <\Phi (t) | i\hbar 
{\partial \over \partial t} - {\hat H} |\Phi (t) > \; , 
\eeq       
where $\Phi$ is the many-body wave-function of the system and 
\beq 
<\Phi (t) |{\hat A} |\Phi (t) > 
= \int \; d^3{\bf r}_1 ... d^3{\bf r}_N\;  
\Phi^*({\bf r}_1,...,{\bf r}_N,t) {\hat A} 
\Phi ({\bf r}_1,...,{\bf r}_N,t) \; , 
\eeq
for any quantum operator ${\hat A}$. 
Looking for stationary points of $S$ with respect 
to variation of the conjugate wave-function $\Phi^*$ gives 
\beq
i\hbar {\partial \over \partial t}\Phi
({\bf r}_1,...,{\bf r}_N,t) = {\hat H}\; 
\Phi({\bf r}_1,...,{\bf r}_N,t) \; , 
\eeq 
which is the many-body time-dependent Schr\"odinger equation. 
\par 
As is well known, except for integrable systems, 
it is impossible to obtain the exact solution 
of the many-body Schr\"odinger equation 
and some approximation must be used. 
Here we discuss the zero-temperature mean-field approximation 
for a system of trapped weakly-interacting 
bosons in the same quantum state, i.e. a Bose-Einstein condensate. 
In this case the Hartree-Fock equations reduce to only one equation, 
the Gross-Pitaevskii equation, which describes the dynamics 
of the condensate. As previously discussed, 
this equation is intensively studied 
because of the recent experimental achievement of Bose-Einstein 
condensation for atomic gasses 
in magnetic traps at very low temperatures.  
\par 
The Hamiltonian operator of a system of $N$ identical 
bosons of mass $m$ is given by 
\beq 
{\hat H}=\sum_{i=1}^N \Big( -{\hbar^2\over 2 m} \nabla_i^2 
+ U({\bf r}_i) \Big) + 
{1\over 2} \sum_{ij=1}^N V({\bf r}_i,{\bf r}_j) \; , 
\eeq  
where $U({\bf r})$ is an external potential and $V({\bf r},{\bf r}')$ 
is the interaction potential. 
In the mean-field approximation the totally symmetric 
many-particle wave-function of the Bose-Einstein condensate reads 
\beq 
\Phi({\bf r}_1,...,{\bf r}_N,t) = \Psi({\bf r}_1,t) 
\Psi({\bf r}_2,t)\; ... \; \Psi({\bf r}_{N-1},t) 
\Psi({\bf r}_N,t) \; , 
\eeq 
where $\Psi ({\bf r},t)$ is the single-particle wave-function. 
Note that such factorization of the total wave-function is exact 
in the case of a non-interacting condensate. 
The quantum action of the system is then simply given by  
\beq 
S_{GP}=N \int dt <\Psi (t) | i\hbar 
{\partial \over \partial t} - {\hat h_s} |\Psi (t) > \; , 
\eeq      
where 
\beq 
{\hat h_s}= -{\hbar^2\over 2 m} \nabla^2 
+ U({\bf r}) + {1\over 2}(N-1) 
\int d^3{\bf r}' |\Psi ({\bf r}',t)|^2 V({\bf r},{\bf r}') \; . 
\eeq  
We call $S_{GP}$ the Gross-Pitaevskii (GP) action of 
the Bose condensate [10]. 
By using the quantum least action principle  
we get the Euler-Lagrange equation 
\beq
i\hbar {\partial \over \partial t}\Psi ({\bf r},t)= 
\Big[ -{\hbar^2\over 2m} \nabla^2 
+ U({\bf r}) + (N-1) 
\int d^3{\bf r}' V({\bf r},{\bf r}') |\Psi ({\bf r}',t)|^2 
\Big] \Psi ({\bf r},t)  \; , 
\eeq 
which is an integro-differential nonlinear Schr\"odinger equation. 
Such equation and the effect of a finite-range interaction have 
been analyzed only by few authors [11,12]. 
In fact, at low energies, it is possible to substitute 
the true interaction with a pseudo-potential  
\beq 
V({\bf r},{\bf r}') = g \; \delta^3 ({\bf r}-{\bf r}') \; , 
\eeq 
where $g={4\pi \hbar^2 a_s/m}$ is the scattering amplitude and $a_s$ 
the s-wave scattering length. The scattering length is positive 
(repulsive interaction) for $^{87}$Rb and $^{23}$Na atoms but 
negative (attractive interaction) for $^7$Li atoms.  
Moreover, for large $N$, the factor $(N-1)$ can be substituted with $N$. 
In this way one obtains the so-called time-dependent GP equation 
\beq 
i\hbar {\partial \over \partial t}\Psi ({\bf r},t)= 
\Big[ -{\hbar^2\over 2m} \nabla^2 
+ U({\bf r}) + g N |\Psi ({\bf r},t)|^2 \Big] \Psi ({\bf r},t)  \; , 
\eeq  
that is the starting point of many calculations [13]. 
Note that the GP equation is accurate to describe the condensate 
of weakly-interacting bosons only 
near zero temperature, where thermal excitations can be neglected. 
The ground-state solution $\psi({\bf r})$ of the GP equation is found 
setting $\Psi({\bf r},t)=e^{-i\mu/\hbar} \psi({\bf r})$ 
in the previous equation. In this way one gets 
the stationary GP equation 
\beq
\Big[ -{\hbar^2\over 2m} \nabla^2 
+ U({\bf r}) + g N |\psi ({\bf r})|^2 \Big] \psi ({\bf r}) 
= \mu \; \psi ({\bf r}) \; , 
\eeq
where $\mu$ is the chemical potential. Note that this equation 
can be also found by mimimizing the GP energy functional 
\beq
E = N \int d^3{\bf r} \;  
{\hbar^2\over 2m} |\nabla \psi ({\bf r})|^2 
+ U({\bf r}) |\psi ({\bf r})|^2 +{1\over 2}g N |\psi ({\bf r})|^4 \; , 
\eeq 
with the normalization condition 
\beq 
\int d^3{\bf r} \; |\psi ({\bf r})|^2 = 1 \; , 
\eeq 
which fixes the chemical potential $\mu$, that is 
the Lagrange multiplier of the stationary minimization problem. 
\par 
To calculate the energy and wavefunction of the zero-temperature 
elementary excitations, 
one must solve the so--called Bogoliubov--de Gennes (BdG) equations [14]. 
The BdG equations can be obtained from the linearized 
time-dependent GP equation. Namely, one can look for zero angular momentum 
solutions of the form 
\beq 
\Psi ({\bf r},t) = e^{-{i\over \hbar}\mu t} 
\Big[ \psi ({\bf r}) + u({\bf r}) e^{-i \omega t} 
+v^*({\bf r}) e^{i \omega t} \Big] \; , 
\eeq 
corresponding to small oscillations of the wavefunction 
around the ground state solution $\psi({\bf r})$. 
By keeping terms linear in the complex functions 
$u({\bf r})$ and $v({\bf r})$, one finds 
the following BdG equations 
$$ 
\Big[ -{\hbar^2\over 2m} \nabla^2 
+ U(\rho,z) - \mu + 2g N |\psi ({\bf r})|^2 \Big] u({\bf r}) 
+ g N |\psi ({\bf r})|^2 v({\bf r}) = \hbar \omega \; u({\bf r}) \; , 
$$ 
\beq 
\Big[ -{\hbar^2\over 2m} \nabla^2 
+ U(\rho,z) - \mu + 2g N |\psi ({\bf r})|^2 \Big] v({\bf r}) 
+ g N |\psi ({\bf r})|^2 u({\bf r})= - \hbar \omega \; v({\bf r}) \; . 
\eeq 
The BdG equations allow one to calculate the eigenfrequencies $\omega$ 
and hence the energies $\hbar \omega$ of the elementary 
excitations. This procedure is equivalent to the 
diagonalization of the N--body Hamiltonian of the system in the Bogoliubov 
approximation. 

\subsection{Bose condensate in a toroidal potential} 

In this section we consider a 3-D toroidal trap given by a quartic 
Mexican hat potential along the cylindrical radius and a harmonic potential 
along the $z$ axis. As we have recently shown [15], 
the resulting toroidal trapping potential is very flexible and it is possible 
to modify considerably the density profile of the BEC 
by changing the parameters of the potential or the number of bosons. 
Moreover, the toroidal trap can be used to create a superfluid, 
namely persistent currents in absence of imposed rotation. 
\par 
The toroidal trap we are discussing can be described  
by the following potential  
\beq 
U({\bf r})={1\over 4} \lambda (\rho^2-\rho_0^2)^2
+{1\over 2} m\omega_z^2 z^2 \; , 
\eeq 
where $\rho=\sqrt{x^2+y^2}$ and $z$ 
are the cylindrical coordinates. 
The potential $U({\bf r})$ is minimum 
along the circle of radius $\rho = \rho_0$ at $z=0$ and 
it has a local maximum at the origin in the $(x,y)$ plane. 
Small oscillations in the $(x,y)$ plane around $\rho_0$ have a 
frequency $\omega_{\perp}=\rho_0 (2\lambda /m)^{1/2}$.  
\par 
In the Thomas-Fermi approximation,  
i.e. neglecting the kinetic energy in the stationary 
GP equation, we find  
\beq 
\psi ({\bf r}) = \Big[ {1\over g N}(\mu - U({\bf r})) \Big]^{1/2} 
\Theta (\mu - U({\bf r}))  \; , 
\eeq 
where $\Theta (x)$ is the step function. For our system 
we obtain that: 
a) the wave function has its maximum value at $\rho =\rho_0$ and $z=0$; 
b) for $\mu < \lambda \rho_0^4/4$ the wave function has a toroidal shape; 
c) for $\mu > \lambda \rho_0^4/4$ the wave function 
has a local minimum at $\rho=z=0$; 
d) the chemical potential scales as $\mu \sim N^{1/2}$.  
It is important to note that the TF approximation neglects 
tunneling effects: to include these processes, it is 
necessary to analyze the full GP problem. 
\par
We have performed the numerical minimization of the GP functional 
by using the steepest descent method. 
This method consists of projecting onto the minimum of the energy 
functional an initial trial state by propagating it in imaginary time. 
We have discretize the space with a grid of points taking advantage of 
the cylindrical symmetry of the problem. 
We have used grids up to $200\times 200$ points verifying that the 
results do not depend on the discretization parameters. 
The number of iterations in imaginary 
time depends on the degree of convergence required and the goodness 
of the initial trial wave function. 
In calculations we have adopted $z$--harmonic oscillator units. 
We write $\rho_0$ in units $a_z=(\hbar / (m \omega_z))^{1/2} =1\;\mu$m, 
$\lambda$ in units $(\hbar \omega_z)a_z^{-4}= 0.477 \; (5.92)$ peV/$\mu$m$^4$ 
and the energy in units $\hbar \omega_z= 0.477 \; (5.92)$ peV 
for $^{87}$Rb ($^{7}$Li). 
Moreover, we have used the following values for the scattering length: 
$a_s=50 \;(-13)\; \AA$ for $^{87}$Rb ($^{7}$Li). 
\par 
In the case of positive scattering length ($^{87}$Rb) 
we can control the density profile of the BEC by modifying 
the parameters of the potential and also the number of particles. 
For small number of particles the condensate is essentially 
confined along the minimum of $U({\bf r})$, there is a very small 
probability of finding particles in the center of the trap so 
that the system is effectively multiply connected. As $N$ increases 
the center of the trap starts to fill up and the system becomes 
simply connected. The value of $N$ for which there is 
a crossover between the two regimes increases with 
the value of $\lambda$ and of $\rho_0$ and, within Thomas-Fermi
approximation,  scales like $\lambda^{3/2}\rho_0^8$. 
\par 
In the case of negative scattering length ($^{7}$Li), 
it is well known that for the BEC in harmonic potential 
there is a critical number of bosons $N_c$, 
beyond which there is the collapse of the wave function 
(see also [16]). 
We obtain the same qualitative behavior for the $^{7}Li$ condensate in our 
Mexican hat potential. However, in cylindrical symmetry, the collapse 
occurs along the line which characterizes the minima of the external
potential, i.e. at $\rho=\rho_0$ and $z=0$. We notice that, 
for a fixed $\rho_0$, the critical number of bosons $N_c$ 
is only weakly dependent on the height of the barrier of the 
Mexican potential. These results suggest that we can not use 
toroidal traps to significantly enhance the 
metastability of the BEC with negative scattering length. 
\par 
It is useful to study states having a vortex line 
along the $z$ axis and all bosons flowing around it with 
quantized circulation. The observation of these vortex 
states is a signature of macroscopic phase coherence of trapped BEC. 
The axially symmetric condensate wave function can be written as 
\beq
\psi_k({\bf r}) = \psi_k (\rho , z) \; e^{i k \theta} \; , 
\eeq
where $\theta$ is the angle around the $z$ axis and $k$ is the integer 
quantum number of circulation. The resulting GP energy functional 
can be written in terms of $\psi_k({\bf r})$ by taking advantage of the 
cylindrical symmetry of the problem: 
$$ 
E = N \int \rho \; d\rho \; dz \; d\theta \; {\hbar^2\over 2m} 
\Big[ 
|{\partial \psi_k (\rho,z)\over \partial\rho}|^2 + 
|{\partial \psi_k (\rho,z)\over \partial z}|^2 \Big]+ 
$$
\beq  
+\Big[{\hbar^2 k^2 \over 2m\rho^2} 
+ U(\rho,z) \Big] \left\vert\psi_k (\rho,z)\right\vert^2 
+{1\over 2}g N|\psi_k (\rho,z)|^4  \; . 
\eeq 
Due to the presence of the centrifugal term, the solution 
of this equation for $k\neq 0$ has to vanish on the $z$ axis 
providing a signature of the vortex state. 
\par 
As previously stated, vortex states are important to characterize the macroscopic 
quantum phase coherence and also superfluid properties 
of Bose systems. It is not difficult to calculate the critical frequency 
$\Omega_c$ at which a vortex can be produced. 
One has to compare the energy of a vortex state in a frame rotating 
with angular frequency $\Omega$, that is $E-\Omega L_z$, with 
the energy of the ground state with no vortices. Since the 
angular momentum per particle is $\hbar k$, the critical 
frequency is given by $\hbar \Omega_c =(E_k/N - E_0/N)/k$, 
where $E_k/N$ is the energy per 
particle of the vortex with quantum number $k$. 
In Table 1 we show some numerical results for vortices of $^{87}Rb$. 
The critical frequency turns out to increase slightly 
with the number of atoms. This corresponds to a moderate 
lowering of the momentum of inertia per unit mass of the condensate 
when $N$ grows. 

\vskip 0.5cm 

\begin{center}
\begin{tabular}{|cccc|} \hline
$N$ & $E_1/N$ & $\mu_1$ & $\hbar \Omega_c$ \\ 
\hline
$5000$  & $6.00$  & $7.87$  & $0.15$  \\ 
$10000$ & $7.61$  & $10.44$ & $0.16$  \\
$20000$ & $10.02$ & $14.22$ & $0.18$  \\
$30000$ & $11.93$ & $17.20$ & $0.20$  \\
$40000$ & $13.57$ & $19.75$ & $0.21$  \\ 
$50000$ & $15.04$ & $22.04$ & $0.23$  \\ 
\hline 
\end{tabular} 
\end{center} 
\vskip 0.3 truecm 
{\bf Table 1}. Toroidal trap. Vortex states of $^{87}Rb$ atoms 
with $k=1$ in the toroidal trap with $\rho_0 =2$ and $\lambda=4$ 
within Hartree-Fock approximation. 
Chemical potential and energy are in units of $\hbar \omega_z=
0.477$ peV ($\omega_z=0.729$ kHz). 
Lengths are in units of $a_z=1\;\mu$m. 

\vskip 0.5cm 

\par 
Once a vortex has been produced, the Bose condensate is superfluid if 
the circulating flow persists, in a metastable state, 
in the absence of an externally imposed rotation. 
Vortex solutions centered in 
harmonic traps have been found, but such states turn out 
to be unstable to single particle excitations out of the condensate. 
The vortex state is superfluid 
if only if its lowest elementary excitation is positive [17]. 
To have the complete spectrum of elementary excitations, 
one must solve the BdG equations (see Eq. (27)). 
We have solved the two BdG eigenvalue equations by finite-difference 
discretization with a lattice of $40\times 40$ points 
in the $(\rho,z)$ plane. In this way, the eigenvalue problem reduces to the 
diagonalization of a $3200\times 3200$ real matrix. 

\vskip 0.5cm 

\begin{center}
\begin{tabular}{|c|cccccc|} \hline
$N$ & $5\times 10^3$ & 
$10^4$ & $2\times 10^4$ & $3\times 10^4$ & $4\times 10^4$ 
& $5\times 10^4$ \\ 
$\hbar\omega$ & $1.22$ & $1.48$ & $1.73$ & $1.88$ & $1.99$ & $2.08$ \\ 
\hline 
\end{tabular}
\end{center} 
\vskip 0.3 truecm 
{\bf Table 2}. Toroidal trap. Bogoliubov elementary excitation 
for a vortex state of $^{87}Rb$ atoms 
with $k=1$ in the toroidal trap with $\rho_0 =2$ and $\lambda=4$. 
Units as in Tab. 1. 

\vskip 0.5cm 

We have tested our program in simple models by comparing numerical
results with the analytical solution and 
verified that a $40 \times 40$ mesh already gives reliable 
results for the lowest part of the spectrum. 
The results are shown in Table 2: 
The lowest Bogoliubov excitation is always positive. 
We have also verified that vortex states become unstable by strongly reducing 
either density (down to about one hundred bosons in our model trap) 
or scattering length. Thus, we can conclude that a 
strongly-interacting Bose condensate 
in a toroidal potential is superfluid.  

\subsection{Bose condensate in a double-well potential}

In a recent experiment at MIT [18], the macroscopic interference 
of two Bose condensates released from the double minimum 
potential has been demonstrated. 
Here, we consider the same double-well potential. 
We analyze the ground state properties of 
the condensate and calculate the spectrum of 
the Bogoliubov elementary excitations as a function of 
the double-well barrier. By varying the strength of 
the barrier one can observe macroscopic quantum effects, 
like the formation of two Bose condensates, 
the collective oscillations and the quantum tunneling [19]. 
\par 
The double-well trap is given by a harmonic anisotropic 
potential plus a Gaussian 
barrier along the $z$ axis, which models the effect of a laser 
beam perpendicular to the long axis of the condensate:  
\beq
U({\bf r})={1\over 2}m\omega_{\rho}^2\rho^2+{1\over 2}m\omega_z^2 z^2 
+ U_0 \exp{ \Big( {-z^2\over 2 \sigma^2} \Big)} \; , 
\eeq 
where $\rho=\sqrt{x^2+y^2}$, $z$ and the angle $\theta$ 
are the cylindrical coordinates. 
The parameter values appropriate for Ref. [18] are 
$\omega_{\rho}=2\pi \times 250$ Hz, $\omega_z=2\pi \times 19$ Hz, 
and $\sigma= 6$ $\mu m$. The anisotropic harmonic trap 
implies a cigar-shaped condensate ($\lambda = \omega_z/\omega_{\rho}=
15/250 <1$), where $z$ is the long axis, and the Gaussian barrier 
of strength $U_0$ creates a double-well potential. 
\par
We have performed the numerical minimization of the GP energy 
functional by using the steepest descent method. 
At each time step the matrix elements entering 
the Hamiltonian are evaluated by means of finite-difference approximants 
using a grid of $200\times 800$ points. 
In our calculations we have used the $z$-harmonic oscillator units. 
For $^{23}$Na atoms, the harmonic length is 
$a_z=(\hbar / (m \omega_z))^{1/2} =4.63\;\mu$m 
and the energy is $\hbar \omega_z = 0.78$ peV. 
Moreover, we have used the following value for the 
scattering length: $a_s=3$ nm [3]. 
Most of our computations has been performed for $N=5\times 10^{6}$ atoms, 
a value typical of the MIT experiment [19]. 

\begin{figure}
\centerline{\psfig{file=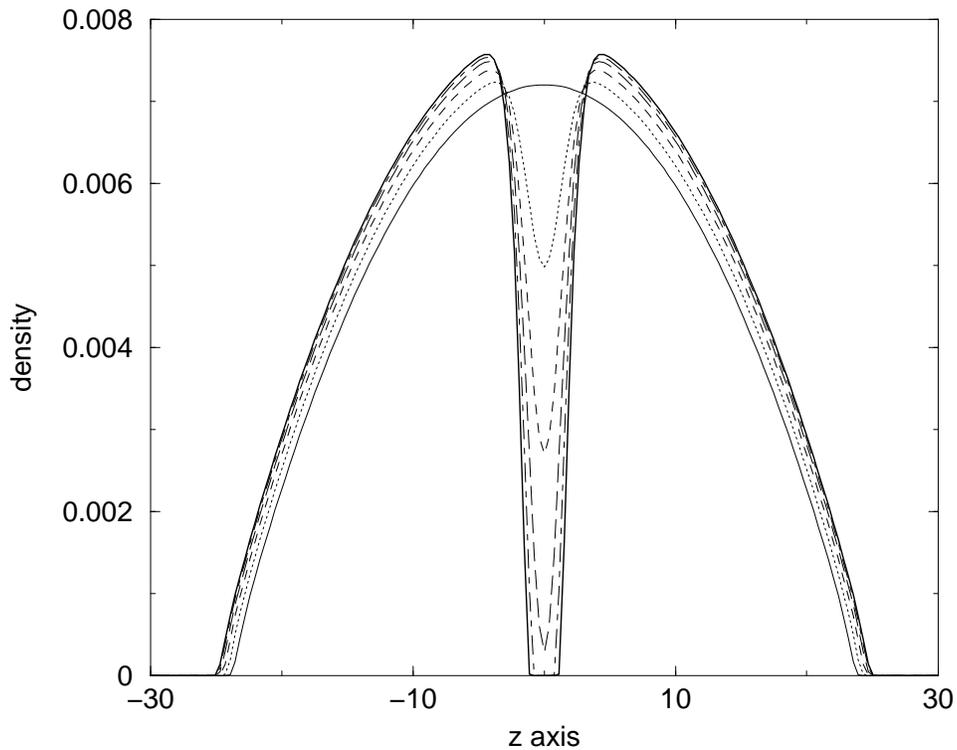,height=4.0in}}
\caption{Double-well trap. Particle probability density 
in the ground state of $N=5\times 10^6$ 
$^{23}$Na atoms as a function of the $z$ axis at $r=0$ (symmetry plane). 
The curves correspond to increasing values of the strength $U_0$ 
of the barrier (from $0$ to $500$), in units of $\hbar \omega_z = 0.78$ peV. 
The laser power is given by the conversion formula 
$P=0.09 \times U_0$ mW. Lengths are in units of $a_z =4.63\; \mu$m.}  
\end{figure}

\par 
In Figure 1 we show the ground state density profile of the $^{23}$Na 
condensate for different values of the strength of the barrier. 
By increasing the strength, the fraction of $^{23}$Na atoms 
decreases in the central region and the Bose condensate separates 
in two condensates. 
The numerically calculated density profiles are in good agreement 
with the phase-contrast images of the MIT experiment [18] 
and with the Thomas-Fermi (TF) approximation, 
which neglects the kinetic term in the GP equation. 
Due to the large number of atoms involved ($N=5\times 10^6$), only 
near the borders of the wave function there are small deviations 
from the TF approximation. 
Note that the potential barrier $U_0$ can be written 
as $U_c/k_B=(37 \mu K)P/\sigma^2$ ($\mu$m$^2$/mW), 
where $P$ is the total power of the 
laser beam perpendicular to the long axis of the condensate and 
$\sigma = 6 \mu$m is the beam radius. 
The conversion factor is $0.09$ mW/($\hbar \omega_z$), 
such that $U_0=100$ (in $\hbar \omega_z$ units) 
gives a laser Power $P=9$ mW [18]. 
\par 
Another important property of the BEC is the spectrum 
of elementary excitations. To calculate the energy and wavefunction 
of elementary excitations, one must solve the BdG equations. 
The excitations can be classified according to 
their parity with respect to the symmetry $z\to -z$. 
We have solved the two BdG eigenvalue equations by finite-difference 
discretization with a lattice of $40\times 40$ points 
in the $(\rho,z)$ plane. In this way, the eigenvalue problem reduces to the 
diagonalization of a $3200\times 3200$ real matrix. 
We have tested our program in simple models by comparing numerical
results with analytical solutions [19]. 

\begin{figure}
\centerline{\psfig{file=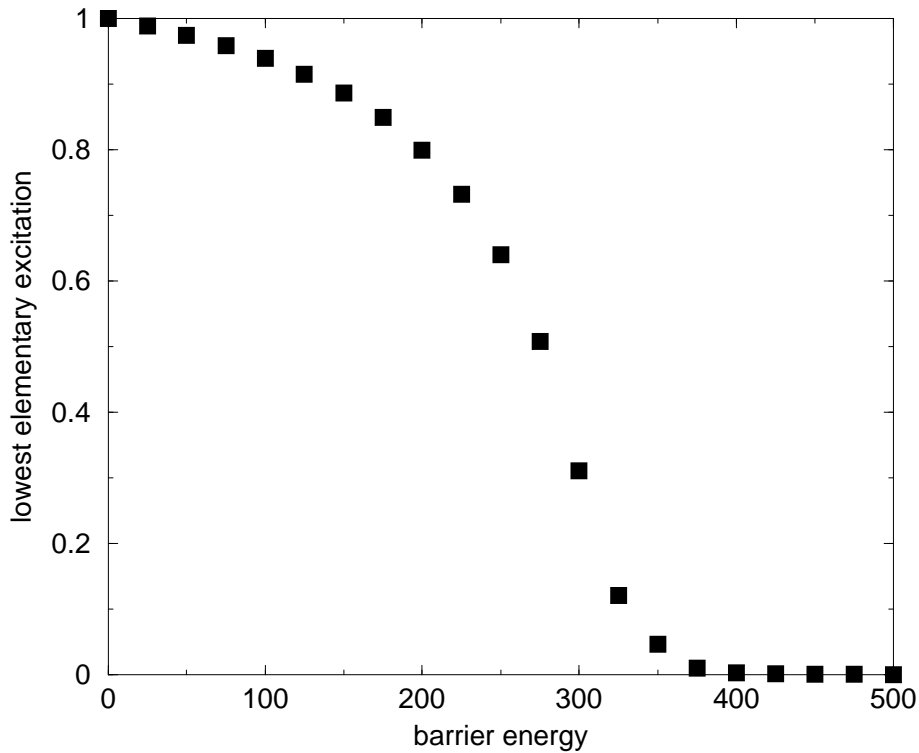,height=4.0in}}
\caption{Double-well trap. 
Lowest elementary excitation $\hbar \omega_1$ 
{\it vs} barrier energy $U_0$ for $N=5\times 10^6$ $^{23}$Na atoms. 
Energies in units of $\hbar \omega_z = 0.78$ peV. 
The laser power is given by the conversion formula 
$P=0.09 \times U_0$ mW.}
\end{figure}

\par 
In Figure 2 we show the lowest elementary excitations of 
the Bogoliubov spectrum for the ground state of the system. 
When the Gaussian barrier is switched off, 
one observes the presence of an odd excitation at energy quite 
close to $\hbar \omega =1$ (in units 
$\hbar \omega_z$). This mode is related to the oscillation of the 
center of mass of the condensate along the z-axis, due to the harmonic 
confinement. This collective oscillation 
is an exact eigenmode of the problem characterized by the 
frequency $\omega_z$, independently of the strength of the interaction. 
The inclusion of the Gaussian barrier modifies the harmonic 
confinement along the $z$-axis 
and this odd collective mode decreases by increasing the strength 
of the barrier. 
\par 
For large values of the Gaussian barrier, i.e. when the 
BEC separates in two condensates, we find quasi-degenerate pairs 
of elementary excitations (even-odd). 
The lowest elementary excitation and the ground-state 
of the GP equation constitute one of such pairs 
and get closer and closer 
as the barrier is increased. This is not surprising 
because in the infinite barrier limit we have two equal 
and independent Bose condensates with the same energy spectrum. 
\par 
An interesting aspect of BEC in double-well traps is the 
possibility to detect the macroscopic quantum tunneling (MQT). 
The MQT has been recently investigated 
by Smerzi et al. [20]. They have found 
that the time-dependent behavior of the condensate in the tunneling 
energy range can be described by the two-mode equations 
\beq
{\dot z}=-\sqrt{1-z^2}\sin{\phi} \; , \;\;\;\;\;\;\; 
{\dot \phi}=\Lambda z +{z\over \sqrt{1-z^2}}\cos{\phi} \; , 
\eeq 
where $z=(N_1-N_2)/N$ is the fractional population 
imbalance of the condensate in the two wells, 
$\phi=\phi_1-\phi_2$ is the relative phase 
(which can be initially zero), and $\Lambda = 4 E^{int}/\Delta E^0$. 
$E^{int}$ is the interaction energy of the condensate 
and $\Delta E^0$ is the kinetic+potential energy 
splitting between the ground state and the quasi-degenerate 
odd first excited state of the GP equation. 
For a fixed $\Lambda$ ($\Lambda >2$), 
there exists a critical $z_c =2\sqrt{\Lambda -1}/\Lambda$ 
such that for $0 < z << z_c$ there are Josephson-like 
oscillations of the condensate with period 
$\tau = \tau_0/\sqrt{1+\Lambda}$, where $\tau_0 =2\pi \hbar /\Delta E^0$. 
But for $z_c<z \leq 1$ there is macroscopic quantum 
self-trapping (MQST) of the condensate: even if the populations 
of the two wells are initially set in an asymmetric state ($z\ne 0$) 
they maintain the original population imbalance without transferring 
particles through the barrier as expected for a free Bose gas. 

\vskip 0.5cm 

\begin{center}
\begin{tabular}{|cccc|} \hline 
$a_s/a_s^{Na}$ & $\Lambda$ & $\tau_0$ (sec) & $z_c$ \\ 
\hline 
$10^{-1}$  & $1108.337$ & $14.583$  & $0.060$ \\ 
$10^{-2}$  & $133.643$  & $13.887$  & $0.173$ \\ 
$10^{-3}$  & $1.390$    & $13.842$  & none    \\ 
$10^{-4}$  & $0.103$    & $10.253$  & none    \\ 
\hline 
\end{tabular} 
\end{center} 

\vskip 0.3 truecm 
{\bf Table 3}. Double-well trap. 
Parameters of the MQT for different 
values of the scattering length $a_s$ with 
$a_s^{Na}=3$ nm and $\tau_0=2\pi \hbar/\Delta E^0$.  
Condensate with $N=5\times 10^3$ atoms. Barrier with 
$U_0=20$ and $\sigma= 1.5$ $\mu$m. 
Energy barrier $U_0$ in units of $\hbar \omega_z = 0.78$ peV. 

\vskip 0.5cm 

\par 
By solving the stationary GP equation 
in the MIT double-well trap with $^{23}$Na, 
we have found that the parameter $\Lambda$ 
is larger than $10^4$ also when few 
particles are present. Nevertheless, one can control 
the dynamics of the condensate by reducing the scattering 
length $a_s$ and the thickness $\sigma$ of the laser beam. 
In particular, as shown in Table 3, 
the parameter $\Lambda$ scales linearly with $a_s$. 
This is an important point because recently it was confirmed 
experimentally the fact that it is now possible to control the 
two-body scattering length by placing atoms in an external 
field. This fact opens the way to a direct observation of 
a macroscopic quantum tunneling of thousands of atoms through 
a potential barrier. 

\section{Conclusions}

We have studied Bose gases in various trapping 
potentials above and below the BEC 
transition temperature. By using the 
semiclassical approximation, we have derived analytical formulas 
for the BEC transition temperature of an ideal Bose gas 
in a generic power-law potential and d-dimensional space. 
Then we have analyzed the effects of the shape of the external 
potential on a zero-temperature Bose condensate by using 
the Gross-Pitaevskii equation, which describes the macroscopic 
wave-function of the condensate, and the Bogoliubov-de Gennnes 
equations, which describe the elementary excitations of the 
condensate. We have shown that, contrary to the 
case of a Bose condensate in a harmonic potential, 
a strongly-interacting condensate in a toroidal potential 
is superfluid.  
Finally, we have investigated the conditions under which 
it is possible to find macroscopic quantum tunneling 
with a Bose condensate in a double-well potential. 

\section*{References}

\begin{description}

\item{\ [1]} M.H. Anderson, J.R. Ensher, M.R. Matthews, C.E. Wieman, 
and E.A. Cornell, Science {\bf 269}, 189 (1995). 

\item{\ [2]} K.B. Davis, M.O. Mewes, M.R. Andrews, N.J. van Druten, 
D.S. Drufee, D.M. Kurn, and W. Ketterle, Phys. Rev. Lett. {\bf 75}, 
3969 (1995).

\item{\ [3]} C.C. Bradley, C.A. Sackett, J.J. Tollet, and R.G. Hulet, 
Phys. Rev. Lett. {\bf 75}, 1687 (1995). 

\item{\ [4]} S. Chu, C. Cohen-Tannouji, and 
W.D. Phillips, Nobel Prize in Physics (1997).  

\item{\ [5]} A. Einstein, 
Preussische Akademie der Wissenshaften 
{\bf 22} 261 (1924); {\bf 1} 3 (1925); {\bf 3} 18 (1925) 

\item{\ [6]} S.N. Bose, Z. Phys. {\bf 26} 178 (1924) 

\item{\ [7]} F. London, Nature {\bf 141} 643 (1938); 
F. London, Phys. Rev. {\bf 54} 947 (1938); 
F. London J. Phys. Chem. {\bf 43} 49 (1938).

\item{\ [8]} K. Huang, {\it Statistical Mechanics} 
(John Wiley, New York, 1987); A.L. Fetter and J.D. Walecka,  
{\it Quantum Theory of Many-Particle Systems} 
(Mc Graw-Hill, Boston, 1971) 

\item{\ [9]} L. Salasnich, J. Math. Phys. {\bf 41} 8016 (2000). 

\item{\ [10]} L. Salasnich, Int. J. Mod. Phys. B {\bf 14} 1 (2000). 

\item{\ [11]} A. Parola, L. Salasnich and L. Reatto, Phys. Rev. A 
{\bf 57}, R3180 (1998); 
L. Reatto, A. Parola and L. Salasnich, J. Low Temp. Phys. 
{\bf 113}, 195 (1998). 

\item{\ [12]} L. Salasnich, Phys. Rev. A {\bf 61}, 015601 (2000). 

\item{\ [13]} E.P. Gross, Nuovo Cimento {\bf 20}, 454 (1961); 
L.P. Pitaevskii, Sov. Phys. JETP {\bf 13}, 451 (1961). 

\item{\ [14]} N.N. Bogoliubov, J. Phys. U.S.S.R. {\bf 11} 23 (1947). 

\item{\ [15]} L. Salasnich, A. Parola and L. Reatto, 
Phys. Rev. A {\bf 59}, 2990 (1999). 

\item{\ [16]} L. Salasnich, Mod. Phys. Lett. B {\bf 11} 1249 (1997); 
L. Salasnich, Mod. Phys. Lett. B {\bf 12}, 649 (1998). 

\item{\ [17]} D.S. Rokhsar, Phys. Rev. Lett. {\bf 79}, 2164 (1997).

\item{\ [18]} M.R. Andrews, C.G. Townsend, H.J. Miesner, 
D.S. Drufee, D.M. Kurn, and W. Ketterle, Science {\bf 275}, 
637 (1997).

\item{\ [19]} L. Salasnich, A. Parola and L. Reatto, 
Phys. Rev. A {\bf 60} 4171 (1999). 

\item{\ [20]} A. Smerzi, S. Fantoni, S. Giovannazzi 
and S.R. Shenoy, Phys. Rev. Lett. {\bf 79}, 4950 (1997). 

\end{description}

\end{document}